# EXCITATION OF THE TRANSITION RADIATION BY A RELATIVISTIC ELECTRON BUNCH AT THE BOUNDARY OF A SEMI-INFINITE PLASMA WAVEGUIDE

*Balakirev V.A., Onishchenko I.N.*
*KIPT, Kharkov, Ukraine,*
*E-mail:onish@kipt.kharkov.ua*

The process of the excitation of an electromagnetic field by a relativistic electron bunch at the input of a semi-infinite plasma waveguide is investigated. The shape and intensity of the short transition electromagnetic pulse are determined and its evolution during propagating in the plasma waveguide is studied. Besides, the influence of the plasma boundary on the spatial structure of plasma wake oscillations in plasma waveguide is considered.
PACS 41.75.Lx, 41.85.Ja, 41.69.Bq

## INTRODUCTION

The processes of excitation of wake Langmuir oscillations in plasma by relativistic electron bunches, as a rule, were considered on the basis of models of unbounded plasma systems [1-9]. The transverse (radial) plasma finiteness is taken into account on the example of plasma waveguides of various geometries in [9-12]. Meanwhile, the presence of longitudinal plasma boundaries can lead to a number of qualitative features of the excitation picture of wake fields in plasma by both relativistic electron bunches and laser pulses [13]. Along with wake plasma oscillations in longitudinally bounded plasma systems, the electron bunch will also excite transition electromagnetic radiation [14-18].

In this work, within the model of a semi-infinite plasma waveguide, we study the effect of a sharp entry plasma boundary on the picture of excitation of wake Langmuir oscillations by a relativistic electron bunch (REB). Attention is also paid to the spatial-temporal structure of the transition electromagnetic pulse and its intensity. It should be noted that the frequency spectrum of the transition electromagnetic pulse contains only frequencies exceeding the plasma frequency. Therefore, the mechanism of transition radiation of a relativistic electron bunch can be an effective tool for obtaining ultra-short electromagnetic pulses in dense plasma [13, 19].

## 1. STATEMENT OF THE PROBLEM. BASIC EQUATIONS

The semi-infinite plasma waveguide $z > 0$ ($z$ is the longitudinal coordinate) is made in the form of a perfectly conducting cylindrical tube of radius $b$ completely filled with a homogeneous plasma. An axisymmetric REB is injected into the plasma from an ideally conducting input end $(z = 0)$. Our problem is to determine the electromagnetic field, including wake Langmuir oscillations, which is excited in a semi-bounded plasma waveguide by a relativistic electron bunch.

The initial system of equations consists of Maxwell's equations

$$rot\vec{E} = -\frac{1}{c}\frac{\partial \vec{H}}{\partial t}, \quad rot\vec{H} = \frac{1}{c}\frac{\partial \vec{D}}{\partial t} + \frac{4\pi}{c}\vec{j}_b, \quad (1)$$

$$div\vec{D} = 4\pi\rho, \quad div\vec{H} = 0,$$

$\rho_b, \vec{j}_b$ are charge density and current of an electron bunch, $\vec{D} = \hat{\varepsilon}\vec{E}$ is electric displacement field, $\hat{\varepsilon}$ is dielectric constant operator of plasma.

The system of Maxwell's equations (1) describes the excitation of an electromagnetic field in plasma by external charges and currents. On the surface of the perfectly conducting end $(z = 0)$ and the side surface, $r = b$ the tangential component of the electric field vanishes

$$\vec{E}_t = 0. \quad (2)$$

We will solve the problem of wake field excitation by an axisymmetric electron bunch in plasma as follows. First, we determine the electromagnetic field $\vec{E}_G, \vec{H}_G$ (Green's function) of a moving charge in the form of an infinitely thin ring with a charge density

$$d\rho = -dQ\frac{1}{v_0}\frac{\delta(r-r_0)}{2\pi r}\delta(t - \frac{z}{v_0} - t_0), \quad (3)$$

where $r$ is radial coordinate, $r_0$ is ring radius, $t_0$ is time of entry of an elementary ring bunch into the waveguide, $v_0$ is bunch velocity, $dQ(r_0, t_0)$ is the elementary charge of the ring related to the current density of the bunch at the entrance to the plasma $(z = 0)$ $j_0(t_0, r_0)$ by the expression

$$dQ = j_0(t_0, r_0)2\pi r_0 dr_0 dt_0.$$

The current density of an elementary ring charge is determined by the expression

$$d\vec{j} = v_0 d\rho \vec{e}_z, \quad (4)$$

$\vec{e}_z$ is unit vector in longitudinal direction.

Let's consider the bunch of electrons with the current density

$$j_0(r_0, t_0) = j_0 R(r_0/r_b) T(t_0/t_b), \quad (5)$$

where the function $R(r_0/r_b)$ describes the dependence of the bunch density on radius (transverse profile), $r_b$ is characteristic transverse bunch size, function $T(t_0/t_b)$ describes the longitudinal density profile of the bunch, $t_b$ is characteristic duration of the bunch. The value $j_0$ is connected with a full charge $Q$ by the relation

$$j_0 = Q/(s_{eff} t_{eff}), \quad (6)$$

where $s_{eff}$ is effective cross section of the bunch

$$s_{eff} = \pi r_b^2 \hat{\sigma}, \quad \hat{\sigma} = 2\int_0^{b/r_b} R(\rho)\rho d\rho,$$

and $t_{eff}$ is effective bunch duration

$$t_{eff} = \hat{\tau} t_b. \quad \hat{\tau} = \int_{-\infty}^{\infty} T(\tau_0) d\tau_0.$$

If we define the electromagnetic field excited by an elementary ring charge and current by the following way

$$\vec{E}_G(r, r_0, z, t - t_0) = dQ\vec{E}(r, r_0, z, t - t_0), \quad (7)$$

where $\vec{E}(r, r_0, z, t - t_0)$ is the electric field excited by a ring bunch with a unit charge, then the full electromagnetic field excited by an electron bunch of finite dimensions is found by summing (integrating) the fields of elementary ring bunches.

$$\vec{E}(r, z, t) = \int_0^b 2\pi r_0 dr_0 \int_{-\infty}^t dt_0 j(r_0, t_0) \vec{E}(r, r_0, z, t - t_0).$$

Taking into account relations (5), (6), this expression can be written as follows

$$\vec{E}(r, z, t) = \frac{2\pi Q}{s_{eff} t_{eff}} \int_0^b R\left(\frac{r_0}{r_b}\right) r_0 dr_0 \times$$

$$\times \int_{-\infty}^t T\left(\frac{t_0}{t_b}\right) \vec{E}(r, r_0, z, t - t_0) dt_0. \quad (8)$$

The next step in solution of the problem is the determination of the electromagnetic field (Green's function) (7) of an elementary ring electron bunch.

## 2. DETERMINATION OF THE GREEN FUNCTION

From the initial system of equations (1) the wave equation for the electric field excited by a thin ring electron bunch with charge (3) and current (4) densities (i.e. Green's function $\vec{E}_G$) follows

$$\hat{\varepsilon}\left(\Delta \vec{E}_G - \frac{1}{c^2}\frac{\partial^2}{\partial t^2}\hat{\varepsilon}\vec{E}_G\right) =$$

$$-\frac{2dQ}{v_0}\left[\frac{v_0}{c^2}\vec{e}_z \frac{\delta(r-r_0)}{r}\hat{\varepsilon}\frac{\partial}{\partial t}\delta(t-t_L) + \right.$$

$$\left. +\vec{\nabla}\left(\frac{\delta(r-r_0)}{r}\delta(t-t_L)\right)\right], \quad t_L = t_0 + \frac{z}{v_0}. \quad (9)$$

Due to the symmetry of the system, the electron bunch will excite electromagnetic waves with field components $E_z, E_r, H_\varphi$. From the vector inhomogeneous wave equation (9) in turn to the equation for the radial component of the electric field in turn to follows

$$\hat{\varepsilon}\left(\frac{1}{r}\frac{\partial}{\partial r}r\frac{\partial}{\partial r}E_{Gr} - \frac{1}{r^2}E_{Gr} + \frac{\partial^2}{\partial z^2}E_{Gr} - \frac{1}{c^2}\frac{\partial^2}{\partial t^2}\hat{\varepsilon}E_{Gr}\right) =$$

$$= -\frac{2dQ}{v_0}\delta(t-t_L)\frac{\partial}{\partial r}\left(\frac{\delta(r-r_0)}{r}\right), \quad (10)$$

Equation (10) must be supplemented with the boundary condition

$$E_{Gr}(z=0) = 0.$$

We represent the component of the electric field $E_{Gr}$ in the form of a series in Bessel functions

$$E_{Gr}(z, r, t) = \sum_{n=1}^{\infty} E_n(z, t) J_1\left(\lambda_n \frac{r}{b}\right),$$

where $\lambda_n$ are the roots of the Bessel function $J_0(x)$. Then, instead of equation (10), we obtain the following equation for the expansion coefficients

$$\hat{\varepsilon}\left(\frac{\partial^2 E_n}{\partial z^2} - \frac{1}{c^2}\frac{\partial^2 \hat{\varepsilon} E_n}{\partial t^2} - \frac{\lambda_n^2}{b^2} E_n\right) = 2Q_n \delta(t-t_L), \quad (11)$$

where

$$Q_n = \frac{dQ}{v_0}\frac{\lambda_n}{bN_n}J_0\left(\lambda_n \frac{r_0}{b}\right), \quad N_n = \frac{b^2}{2}J_1^2(\lambda_n).$$

We represent the expansion coefficients $E_n$ in the form of Fourier integrals with respect to time

$$E_n(z,t) = \int_{-\infty}^{\infty} E_{n\omega}(z) e^{-i\omega t} d\omega,$$

where

$$E_{n\omega}(z) = \frac{1}{2\pi}\int_{-\infty}^{\infty} E_n(z,t) e^{i\omega t} dt$$

is Fourier amplitude. Equation (11) implies the ordinary differential equation for the Fourier component of the radial electric field

$$\frac{d^2}{dz^2}E_{n\omega} + k_z^2(\omega)E_{n\omega} = \frac{Q_n}{\pi \varepsilon(\omega)} e^{i\omega t_L},$$

$$k_z^2(\omega) = k_0^2 \varepsilon(\omega) - \frac{\lambda_n^2}{b^2}, \quad \varepsilon(\omega) = 1 - \frac{\omega_p^2}{\omega(\omega+i\nu)}$$

is the plasma dielectric constant; $\omega_p$ is the Langmuir frequency of plasma; $\nu$ is the effective frequency of collisions. The solution to this equation, which satisfies the boundary condition at the perfectly conducting input end of the waveguide $E_{n\omega}(z=0) = 0$ and the radiation condition at infinity ($z \to \infty$), has the form

$$E_{n\omega}(z) = \frac{Q_n e^{i\omega t_0}}{\pi \varepsilon(\omega)(k_n^2 - k_l^2)}\left[e^{ik_l z} - e^{ik_n z}\right],$$

where $k_l = \omega/v_0$. Accordingly, the expression for the expansion coefficients $E_n(z,t)$ of the electric field can be written as follows

$$E_n(z,t) = E_n^{(inf)}(z,t) - E_n^{(tran)}(z,t), \quad (12)$$

$$E_n^{(inf)}(z,t) = \frac{Q_n}{\pi}\int_{-\infty}^{\infty}\frac{e^{-i\omega(t-t_L)}d\omega}{\varepsilon(\omega)(k_n^2 - k_l^2)}. \quad (13)$$

$$E_n^{(tran)}(z,t) = \frac{Q_n}{\pi}\int_{-\infty}^{\infty}\frac{e^{-i(\omega(t-t_0)-k_n z)}d\omega}{\varepsilon(\omega)(k_n^2 - k_l^2)}. \quad (14)$$

The term $E_n^{(inf)}(z,t)$ in expression (12) describes the electromagnetic field of a ring electron bunch in an unbounded plasma waveguide. In turn, the term $E_n^{(tran)}(z,t)$ describes the field that has arisen due to the presence of the input longitudinal plasma boundary (transition electromagnetic field). The integrand for the expansion coefficient $E_n^{(inf)}(z,t)$ contains only simple poles

$$\omega = \pm \omega_p - i\nu/2, \quad (15)$$

which are located in the lower half-plane of the complex variable near the real axis, as well as the poles located on the imaginary axis

$$\omega = \pm i\omega_{nst}, \qquad (16)$$

where

$$\omega_{nst} = \beta_0 \gamma_0 \omega_{nc}, \; \omega_{nc} = \sqrt{\omega_p^2 + \omega_n^2}, \; \omega_n^2 = \frac{\lambda_n^2 c^2}{b^2},$$

$$\beta_0 = v_0/c, \; \gamma_0 = 1/\sqrt{1-\beta_0^2}.$$

We note what $\omega_{nc}$ is the cutoff frequency in the plasma waveguide. Taking into account the weak dissipation of energy in the plasma gives a rule for bypassing the singular points and we will not take it into account in the final formulas. Calculating the residues at the poles, (15), (16), we find the expression for the expansion coefficients $E_n^{(inf)}(z,t)$ and, ultimately, the expression for the radial component of the electric field $E_{Gr}^{(inf)}(r,z,t)$ of a ring electron bunch in the infinite plasma waveguide

$$E_{Gr}^{(inf)} = E_{Grst}^{(inf)} - \frac{\partial}{\partial r}\Phi_{Gw}^{(inf)}, \qquad (17)$$

where

$$E_{Grst}^{(inf)} = -2dQ\sum_{n=1}^{\infty}\frac{\omega_{nst}}{v_0}F_n^{(1)}(r,r_0)e^{-\omega_{nst}|t-t_L|} \qquad (18)$$

$$\Phi_{Gw}^{(inf)} = 4dQk_p F^{(0)}(r,r_0)\vartheta(t-t_L)\sin\omega_p(t-t_L), \; (19)$$

$$F^{(0)}(r,r_0) = \sum_{n=1}^{\infty} F_n^{(0)}(r,r_0), \qquad (20)$$

$$F_n^{(0)}(r,r_0) = \frac{J_0\left(\frac{\lambda_n}{b}r_0\right)J_0\left(\frac{\lambda_n}{b}r\right)}{J_1^2(\lambda_n)\left(\lambda_n^2 + k_p^2 b^2\right)},$$

$$F^{(1)}(r,r_0) = \frac{\partial}{\partial r}F^{(0)}(r,r_0),$$

$\vartheta(t-t_L)$ is the unit Heaviside function. The first term in (17) describes the quasi-static electromagnetic field of a ring electron bunch moving uniformly and rectilinearly in an unbounded plasma waveguide. The second term in the expression for the radial component of the electric field (17) corresponds to the purely potential wake field of Langmuir oscillations. Series (20) can be summed. As a result, we obtain the following expression for the wake electric potential [9]

$$\Phi_{Gw}^{(inf)} = 2dQk_p\Gamma_0(k_p r, k_p r_0)\vartheta(t-t_L)\sin\omega_p(t-t_L), \; (21)$$

$$\Gamma_0(k_p r, k_p r_0)\begin{cases} I_0(k_p r_0)\Delta_0(k_p b, k_p r), \; r \geq r_0, \\ I_0(k_p r)\Delta_0(k_p b, k_p r_0), \; r \leq r_0, \end{cases}$$

$$\Delta_0(k_p b, k_p r) = I_0(k_p b)K_0(k_p r) - I_0(k_p r)K_0(k_p b).$$

We now turn to the study of the transition electromagnetic field excited in a semi-infinite plasma waveguide by a ring electron bunch. In fact, the problem is reduced to the analysis of the Fourier integral (14), which determines the dependence of the transition electromagnetic field on the longitudinal coordinate and time

$$E_{Gr}^{(tran)}(z,r,t) = \sum_{n=1}^{\infty} E_n^{(tran)}(z,t)J_1\left(\frac{\lambda_n}{b}r\right). \qquad (22)$$

Besides the simple poles (15), (16), the integrand in (14) has branch points $\omega = \omega_{bp}$ and $\omega = -\omega_{bp}^*$, which are determined from the equation $k_n(\omega) = 0$, $\omega_{bp} = \omega_{nc} - iv\omega_p^2/2\omega_{nc}^2$.

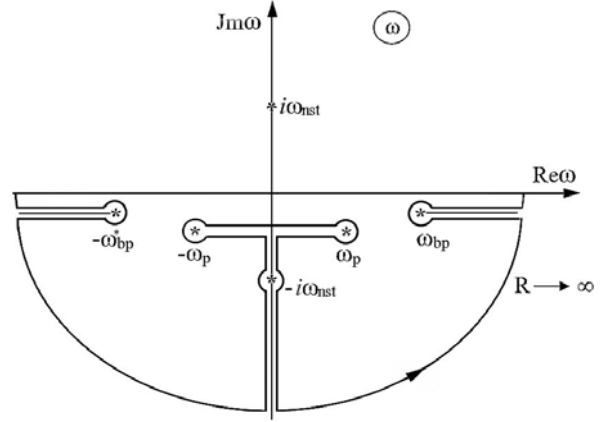

*Fig.1. Integration contour*

The branch points are in the lower half-plane of the complex variable $\omega$. Let's make cuts on the complex plane $\text{Im}\,k_n(\omega) = 0$ (see Fig.1). On the up sheet $\text{Im}\,k_n > 0$ of a two-sheet Riemann surface, the real part $k_n(\omega)$ on the upper and lower sides of the cut have opposite signs. In the upper half-plane $\omega$, the integrand in (14) has only one simple pole $\omega = i\omega_{nst}$. This pole corresponds to the quasi-static (quasi-Coulomb) field of the electron bunch. When

$$z > c(t-t_0) \qquad (23)$$

it is necessary to close the integration contour into the upper half-plane of the complex variable $\omega$. Calculating the residue in the pole $\omega = i\omega_{nst}$, we find the value of the Fourier integral (14) under condition (23)

$$E_n^{(tran)}(z > c(t-t_0)) = E_n^{(inf)}.$$

Accordingly, the full field behind the wave front $z > c(t-t_0)$ is identically zero

$$\vec{E}_G(z > c(t-t_L)) \equiv 0.$$

The result is obvious, since electromagnetic radiation has simply not yet reached this region.

Consider now the region $z < c(t-t_0)$. For this region, the contour of integration in the Fourier integral (14) must be closed in the lower half-plane of the complex variable $\omega$. The selected integration contour is shown in Fig.1. In the inner region bounded by the contour, the integrand in the integral has no singularities. Therefore, the integral along this contour is equal to zero. As a result, we find that the sought Fourier integral (14) is equal to the sum of the residues at the poles (15) corresponding to the Langmuir oscillations, the residue in the pole $\omega = -i\omega_{nst}$ responsible for the quasi-static field of the bunch, plus the integral over the loop $C_{loop}$ encircling the cut. Accordingly, for the expansion coefficients of the transition electric field (14) we have the following expression

$$E_n^{(tran)}(z,\bar{t}) = -2\pi i Q_n \left[ resF_T(\bar{t},z,\omega_p) + resF_T(\bar{t},z,-\omega_p) + \right.$$
$$\left. + resF_T(\bar{t},z,-i\omega_{nst}) \right] + E_n^{(rad)}(z,\bar{t}), \quad (24)$$

where $F_T(\bar{t},z,\omega)$ is the integrand in the integral (14),

$$E_n^{(rad)}(z,\bar{t}) = \frac{Q_n}{\pi} \int_{C_{loop}} \frac{e^{-i(\omega\bar{t}-k_n z)}}{\varepsilon(\omega)(k_n^2 - k_l^2)} d\omega + c.c., \quad (25)$$

$\bar{t} = t - t_0$.

The integral over the cut in the left half-plane of the complex variable $\omega$ ($\text{Re}\,\omega < 0$) is complex conjugate. Integral (25) describes the field of the radiated transition electromagnetic pulse.

Having calculated the residues in the poles, $\omega = \pm\omega_p$ we find expressions for the expansion coefficients and, accordingly, an expression for the electric potential of the transition Langmuir oscillations

$$\Phi_w^{(tran)} = -4dQk_p G_0(r,r_0,z)\vartheta(\bar{t}-z/c)\sin\omega_p\bar{t}. \quad (26)$$

$$G_0(r,r_0,z) = \sum_{n=1}^{\infty} F_n^{(0)}(r,r_0)e^{-\lambda_n \frac{z}{b}}.$$

For the full electric potential of the Langmuir oscillations $\Phi_{Gw}$, we have the expression

$$\Phi_{Gw} = \Phi_w^{(inf)} + \Phi_w^{(tran)} = dQ\Psi(r,r_0,z,\bar{t}), \quad (27)$$
$$\Psi = 4k_p \left[ F^{(0)}(r,r_0)\vartheta(\bar{t}-z/v_0)\sin\omega_p(\bar{t}-z/v_0) - \right.$$
$$\left. - G_0(r,r_0,z)\vartheta(\bar{t}-z/c)\sin\omega_p\bar{t} \right].$$

It is easy to make sure that the electric potential $\Phi_{Gw}$ satisfies the boundary condition

$$\Phi_{Gw}(r,r_0,\bar{t},z=0) = 0.$$

Thus, in a semi-bounded plasma waveguide, the field of Langmuir oscillations excited by an electron bunch contains a monochromatic wake wave, as well as clot of Langmuir oscillations localized near the entrance end of the plasma waveguide. The function $G_0(r,r_0,z)$ describes the spatial distribution of the electric potential of the transition field of Langmuir oscillations.

Let us now consider the structure of a quasi-static transition electromagnetic field of a relativistic electron bunch moving in a semi-infinite plasma waveguide. The residue in the pole $\omega = -i\omega_{nst}$ of the integrand (14) gives an expression for the expansion coefficients in the Bessel series of the radial component of the transient quasi-static electric field of the electron bunch. As a result, we obtain the following final expression for the component of the transition quasi-static field of the electron bunch

$$E_{Grst}^{(tran)} = 2dQ\vartheta\left(\bar{t} - \frac{z}{c}\right)\sum_{n=1}^{\infty}\frac{\omega_{nst}}{v_0}F_n^{(1)}(r,r_0)e^{-\omega_{nst}(\bar{t}+z/v_0)}.$$

This expression describes the field of a positive charge induced at a perfectly conducting plasma boundary by an electron bunch (image charge field). The full quasi-static field of a moving electron bunch will consist of the sum of the fields of the charge itself and its oppositely charged image

$$\vec{E}_{Gst}(r,r_0,z,\bar{t}) = \vec{E}_{Gst}^{(inf)}(r,r_0,|z-v_0\bar{t}|) +$$
$$+ \vec{E}_{Gst}^{(tran)}(r,r_0,z+v_0\bar{t}). \quad (28)$$

At the perfectly conducting plasma boundary $z = 0$, the radial component of the full quasi-static electric field is zero.

Let us proceed to the analysis of the last integral term in expression (24), which describes the field of a transition electromagnetic pulse propagating in a dispersive medium, in our case, in plasma. Taking into account that on the upper and lower sides of the cut (see Fig. 1) $\text{Re}\,k_n(\omega)$ has opposite signs, integral (25) over the loop $C_{loop}$ covering the cut can be transformed to the form

$$E_n^{(rad)}(z,\bar{t}) = -\frac{Q_n}{\pi}v_0^2\gamma_0^2\left[F_n^{(+)}(z,\bar{t}) - F_n^{(-)}(z,\bar{t})\right] + +c.c.,$$

$$F_n^{(\pm)}(z,\bar{t}) = \int_{\omega_{nc}}^{\infty} F(\omega)e^{-i\bar{t}\psi_n^{(\pm)}(\omega)}d\omega, \quad (29)$$

$$F_n(\omega,) = \frac{1}{\varepsilon(\omega)(\omega^2 + \omega_{nst}^2)},$$

$$\psi_n^{(\pm)}(\omega) = \omega \mp \frac{u}{c}(\omega^2 - \omega_{nc}^2)^{1/2}, \, u = \frac{z}{t}.$$

We note that at the perfectly conducting end $z = 0$ of the waveguide, the radial component of the transient electromagnetic pulse also is zero.

Let us change the variable $\kappa = k_n(\omega) \equiv \sqrt{\omega^2 - \omega_{nc}^2}/c$ in integrals (29). We have chosen the longitudinal wave number as the independent variable. In terms of the new variable, integrals (29) take the form

$$F_n^{(\pm)}(z,\bar{t}) = \frac{c}{\omega_{nc}^2\gamma_0^2}\int_0^{\infty} F(\kappa)e^{-i\bar{t}\psi_n^{(\pm)}(\kappa)}d\kappa, \quad (30)$$

where

$$F_n(\kappa) = \frac{\kappa}{\varepsilon(\kappa)\left(1 + \dfrac{\kappa^2}{\gamma_0^2 k_{nc}^2}\right)\sqrt{\kappa^2 + k_{nc}^2}},$$

$$k_{nc} = \frac{\omega_{nc}}{c}, \, \varepsilon(\kappa) = 1 - \frac{\omega_p^2}{\omega^2(\kappa)}, \, \omega(\kappa) = \sqrt{\omega_{nc}^2 + \kappa^2 c^2},$$

$$\psi_n^{(\pm)}(\kappa) = \omega(\kappa) \mp \kappa u, \, u = z/\bar{t}.$$

Let us consider the asymptotic representations of integrals (30) for large $\bar{t}$ [20-22]. In this case, the integrands in (30) oscillate rapidly, and these oscillations compensate each other in most of the integration region. The exceptions are the end point $\kappa = 0$, as well as the points of the stationary phase, in the vicinity of which the phase functions change slowly. The points of stationary phases are the roots of the equations

$$\frac{d\psi_n^{(\pm)}(\kappa)}{d\kappa} = 0 \quad (31)$$

and have the values

$$\kappa = \kappa_{ns}^{(\pm)}(\varsigma) \equiv \pm\frac{k_{nc}\varsigma}{\sqrt{1-\varsigma^2}}, \, \varsigma = \frac{u}{c} \leq 1. \quad (32)$$

Formal equation (31) is equivalent to the physically more visual

$$v_g(\kappa) = \frac{d\omega(\kappa)}{d\kappa} = \pm\frac{z}{t} \equiv \pm c\varsigma.$$

The signs of the group velocity $v_g(\kappa)$ correspond to wave packets propagating in opposite directions. The stationary point $\kappa = \kappa_{ns}^{(+)}(\varsigma)$ of the phase function $\psi_n^{(+)}(\kappa)$ is within the integration of the integral $F_n^{(+)}(z,\bar{t})$ and must be taken into account. The stationary point $\kappa = \kappa_{ns}^{(-)}(\varsigma)$ is outside the limits of integration and does not contribute in the integral $F_n^{(-)}(z,\bar{t})$. In the vicinity of the stationary point $\kappa = \kappa_{ns}^{(+)}(\varsigma)$, the phase function $\psi_n^{(+)}(\kappa)$ can be expanded in a series

$$\psi_n^{(+)}(\kappa) = \psi_n^{(+)}(\kappa_{ns}^{(+)}) + \frac{1}{2}\omega''(\kappa_{ns}^{(+)})\left(\kappa - \kappa_{ns}^{(+)}\right)^2,$$

where

$$\omega''(\kappa_{ns}^{(+)}(\varsigma)) = \frac{c^2}{\omega_{nc}}\left(1-\varsigma^2\right)^{3/2},$$

$$\psi_n^{(+)}(\kappa_{ns}^{(+)}) = \omega(\kappa_{ns}^{(+)}(\varsigma)) - \kappa_{ns}^{(+)}(\varsigma)c\varsigma.$$

The dependence of the longitudinal wave number $\kappa_{ns}^{(+)}$ on the parameter $\varsigma$ is determined by expression (32). For the frequency $\omega_{ns}(\varsigma) \equiv \omega(\kappa_{ns}^{(+)}(\varsigma))$ and dielectric constant $\varepsilon_n(\varsigma) \equiv \varepsilon(\omega_{ns}(\varsigma))$, we have the simple expressions

$$\omega_{ns}(\varsigma) = \frac{\omega_{nc}}{\sqrt{1-\varsigma^2}}, \quad \varepsilon_n(\varsigma) = \frac{1+\sigma_n^2\varsigma^2}{1+\sigma_n^2}, \quad \sigma_n^2 = \frac{\omega_p^2}{\omega_n^2}. \quad (33)$$

The dielectric constant of the plasma varies from its value at the cutoff frequency $\varepsilon(\omega_{nc}) = 1 - \omega_p^2/\omega_{nc}^2$ at the plasma boundary $z = 0$ to unity at the front of the transition pulse $z = c\bar{t}$. If the condition

$$\Lambda = \frac{1}{2}\bar{t}\,\omega''(\kappa_{ns}^{(+)})\kappa_{ns}^{(+)2} = \frac{1}{2}\omega_{nc}\bar{t}\varsigma^2\sqrt{1-\varsigma^2} \gg 1 \quad (34)$$

is satisfied then the narrow vicinity of the stationary phase point $\kappa = \kappa_{ns}^{(+)}(\varsigma)$ gives the main contribution to the integral $F_n^{(+)}(z,\bar{t})$. In this limiting case, we obtain the following approximate value of the integral

$$F_n^{(+)}(\bar{t},\varsigma) = \frac{c}{\omega_{nc}^2\gamma_0^2}F(\kappa_{ns}^{(+)})\sqrt{\frac{2\pi}{\bar{t}\,\omega_{ns}''(\varsigma)}} \times$$

$$\times e^{-i\left[\omega_{ns}(\varsigma)\bar{t} - \kappa_{ns}^{(+)}(\varsigma)z + \frac{\pi}{4}\right]}, \quad (35)$$

where $\omega_{ns}''(\varsigma) \equiv \omega''(\kappa_{ns}^{(+)}(\varsigma))$.

As noted above, the stationary phase point $\kappa = \kappa_{ns}^{(-)}(\varsigma)$ is outside the integration region of the integral $F_n^{(-)}(z,\bar{t})$. Under these conditions, at $\bar{t} \gg 1$, only the end point $\kappa = 0$ will make the main contribution to the value of this integral. In principle, this end point will give a certain contribution to the integral $F_n^{(+)}(z,\bar{t})$ too. The phase functions $\psi_n^{(\pm)}(\kappa)$ and all its derivatives are continuous in the region of integration; therefore, the contribution of the end point to the asymptotic estimation of the integrals under $\bar{t} \to \infty$ can be obtained by using integration by parts [21, 22]. As a result, we obtain the following expressions for the functions $F_{end}^{(\pm)}(\bar{t},\varsigma)$ which take into account the contribution of the end point $\kappa = 0$ to integrals (30)

$$F_{end}^{(\pm)}(\bar{t},z) = -\frac{c}{z^2}F(0)e^{-i\omega_{nc}\bar{t}}.$$

From expression (28) for the electric field of the transition electromagnetic pulse, it is seen that the terms $F_{end}^{(\pm)}(\bar{t},z)$ compensate each other and, accordingly, do not contribute to the full asymptotic expression for coefficient (28). As a result, for the amplitudes of the radial harmonics of the transition electromagnetic radiation in the semi-infinite plasma waveguide, we obtain the following expression

$$E_n^{(rad)}(z,\bar{t}) = -2\frac{dQ}{b^2}X_n(r_0)\frac{\beta_0 c^2}{\omega_{nc}^2 b}F(\kappa_{ns}^{(+)})\sqrt{\frac{2}{\pi\bar{t}\,\omega_{ns}''(\varsigma)}} \times$$

$$\times \cos\left[\omega_{ns}(\varsigma)\bar{t} - \kappa_{ns}^{(+)}(\varsigma)z - \frac{\pi}{4}\right], \quad (36)$$

where $X_n(r_0) = \dfrac{\lambda_n J_0(\lambda_n r_0/b)}{J_1^2(\lambda_n)}$.

Thus, a transition electromagnetic pulse is a wave packet with variable frequency $\omega_{ns}(\varsigma)$ and wave number $\kappa_{ns}^{(+)}(\varsigma)$. It follows from relations (33), (32), that the frequency $\omega_{ns}(\varsigma)$ and local wave number $\kappa_{ns}^{(+)}(\varsigma)$ move in space with the group velocity $v_g(\kappa_{ns}^{(+)}) = c\varsigma = z/\bar{t}$, i.e. conserve its values along the characteristic $z = v_g(\kappa_{ns}^{(+)})\bar{t}$. Note that the group and phase $v_{ph}(\varsigma) = c/\varsigma$ velocities are related by a well-known relationship $v_g(\varsigma)v_{ph}(\varsigma) = c^2$. It also follows from the expression for frequency (33) and wave number (32) that the high-frequency (short-wave) components of the transition electromagnetic pulse is concentrated in the region of the leading edge of the pulse $z \leq c\bar{t}$. With distance from the leading edge to the end of the waveguide, the frequency decreases, and the wavelength, respectively, increases.

Using explicit expressions for, $\omega_{ns}(\varsigma)$ $\kappa_{ns}^{(+)}(\varsigma)$ and $\omega_{ns}''(\varsigma)$, expression (36) for the amplitudes of the radial harmonics of the transition pulse can be written in the form

$$E_n^{(rad)}(z,\bar{t}) = -2\frac{dQ}{b^2}\frac{v_0}{\omega_{nc}b}f(\varsigma)\frac{X_n(r_0)}{\varepsilon_n(\varsigma)}\sqrt{\frac{2}{\pi\omega_{nc}\bar{t}}} \times$$

$$\times \cos\left(\omega_{nc}\bar{t}\sqrt{1-\varsigma^2} - \frac{\pi}{4}\right),$$

where

$$f(\varsigma) = \frac{\varsigma\left(1-\varsigma^2\right)^{1/4}}{1-\beta_0^2\varsigma^2}.$$

The full field of a transition electromagnetic pulse is the superposition of radial harmonics

$$E_{Gr}^{(rad)}(z,r,\bar{t}) = -2\frac{dQ}{b^2}\frac{v_0}{\omega_p b}\sqrt{\frac{2}{\pi\omega_p\bar{t}}}f(\varsigma)\sum_{n=1}^{\infty}\frac{X_n}{\varepsilon_n(\varsigma)}\left(\frac{\omega_p}{\omega_{nc}}\right)^{3/2} \times$$

$$\times J_1\left(\lambda_n\frac{r}{b}\right)\cos\left(\omega_{nc}\bar{t}\sqrt{1-\varsigma^2} - \frac{\pi}{4}\right). \quad (37)$$

Let us also give a similar expression for the magnetic field of the electromagnetic pulse

$$H_{G\varphi}^{(rad)}(z,r,\bar{t}) = -2\frac{dQ}{b^2}\frac{v_0}{\omega_p b}\sqrt{\frac{2}{\pi\omega_p\bar{t}}}\frac{f(\varsigma)}{\varsigma}\sum_{n=1}^{\infty}X_n(r)\left(\frac{\omega_p}{\omega_{nc}}\right)^{3/2} \times$$

$$\times J_1\left(\lambda_n\frac{r_0}{b}\right)\cos\left(\omega_{nc}\bar{t}\sqrt{1-\varsigma^2}-\frac{\pi}{4}\right). \quad (38)$$

In the considered case of a ultra-relativistic electron bunch $\gamma_0 \gg 1$, the function $f(\varsigma)$ that determines the spatial structure of the transition pulse at each instant of time has a sharp maximum at the point

$$\varsigma = \varsigma_m = 1 - \frac{1}{6\gamma_0^2}. \quad (39)$$

The function $f(\varsigma)$ at this point reaches its maximum value

$$f(\varsigma_m) = \frac{3^{3/4}}{4}\gamma_0^{3/2} \gg 1.$$

As a whole, the picture of the evolution of a transition electromagnetic pulse during its propagation in a plasma waveguide is as follows. A sharp maximum of the amplitude of the wave packet is located in the vicinity of the leading edge of the pulse

$$z_m = v_m\bar{t}, \quad v_m = c\left(1 - \frac{1}{6\gamma_0^2}\right)$$

and propagates with a velocity closed to the speed of light. The frequency (33) of the wave packet at this point is equal

$$\omega_{ns}(\varsigma = \varsigma_m) = \sqrt{3}\gamma_0\omega_n$$

and in the considered case of a relativistic electron bunch $\gamma_0 \gg 1$ significantly one exceeds the cutoff frequency and, accordingly, the plasma frequency. Since the frequency of the wave packet and the group velocity decrease with distance from the leading edge, the low-frequency components of the wave packet are constantly lagging behind. As a result, in the process of propagation the width of the wave packet increases and its amplitude decreases as $1/\sqrt{\bar{t}}$, i.e. dispersion spreading of the transition pulse takes place.

## 3. TRANSITION ELECTROMAGNETIC FIELD OF AN ELECTRON BUNCH OF FINITE SIZES

As noted above, in a semi-infinite plasma waveguide the electron bunch excites potential plasma oscillations at the electron Langmuir frequency, its own quasi-static electromagnetic field and the field of its positively charged image, as well as a transition electromagnetic pulse.

Let us first consider the process of excitation of potential plasma oscillations by an electron bunch of finite sizes. We will consider a symmetric electron bunch $T(t_0) = T(-t_0)$, the density of which has a maximum at the time of enter $t_0 = 0$ ($T(0) = 1$) and decreases with distance from it $T(t_0 \to \pm\infty) \to 0$. The electric potential of plasma oscillations excited by an elementary ring electron bunch (Green's function) is described by formula (27). Then, for an electron bunch with profile (5), we obtain the following expression

$$\Phi_w(r,z,t) = \Phi_w^{(ins)}(r,t-z/v_0) + \Phi_w^{(tran)}(r,z,t), \quad (40)$$

where

$$\Phi_w^{(ins)} = \frac{4Qk_p}{\hat{\sigma}\hat{\tau}t_b}\Pi_0(r)\int_{-\infty}^{t-\frac{z}{v_0}}T\left(\frac{t_0}{t_b}\right)\sin\omega_p\left(t-\frac{z}{v_0}-t_0\right)dt_0 \quad (41)$$

is the electric potential of a wake plasma wave in an infinite plasma waveguide,

$$\Pi_0(r) = \sum_{n=1}^{\infty}\frac{Y_n}{J_1^2(\lambda_n)(\lambda_n^2 + k_p^2b^2)}J_0\left(\lambda_n\frac{r}{b}\right),$$

$$Y_n = 2\int_0^{1/\rho_b}R(\eta_0)J_0(\lambda_n\rho_b\eta_0)\eta_0 d\eta_0,$$

$$\eta_0 = r_0/r_b, \quad \rho_b = r_b/b,$$

$$\Phi_w^{(tran)} = -\frac{4Qk_p}{\hat{\sigma}\hat{\tau}t_b}\Pi(r,z)\int_{-\infty}^{t-z/c}T\left(\frac{t_0}{t_b}\right)\sin\omega_p(t-t_0)dt_0 \quad (42)$$

is electric potential of the near-wall clot of transition plasma oscillations,

$$\Pi(r,z) = \sum_{n=1}^{\infty}\frac{Y_n J_0(\lambda_n r/b)}{J_1^2(\lambda_n)(\lambda_n^2 + k_p^2b^2)}\exp(-\lambda_n z/b)$$

is the function that describes the spatial distribution of transition plasma oscillations. We note that

$$\Pi_0(r) = \Pi(r,z=0).$$

At large distances behind the bunch (in the wave zone) $\tau \gg t_b$, the upper limits in integrals (41), (42) can be replaced by infinity. Then for the field of plasma oscillations we obtain

$$\Phi_w(r,z,t) = \frac{4\pi Qk_p}{\hat{\sigma}\hat{\tau}}\hat{T}(\Omega_p)\big[\Pi_0(r)\sin\omega_p(t-z/v_0) - \Pi(r,z)\sin\omega_p t\big],$$

where

$$\hat{T}(\Omega_p) = 2\int_0^{\infty}T(\tau_0)\cos\Omega_p\tau_0 d\tau_0$$

is the amplitude of the Fourier component at the dimensionless plasma frequency $\Omega_p = \omega_p t_b$ of the function $T(t_0/t_b)$ describing the longitudinal profile of the bunch.

We consider an electron bunch with a Gaussian longitudinal profile

$$T\left(\frac{t_0}{t_b}\right) = \exp\left(-\frac{t_0^2}{t_b^2}\right). \quad (43)$$

and the following model transverse particle density profile

$$R(\eta_0) = \begin{cases}\dfrac{J_0(\lambda_{1,1}\eta_0) + s_0}{1+s_0}, & \eta_0 < 1, \\ 0, & \eta_0 > 1,\end{cases} \quad (44)$$

where $\eta_0 = r_0/r_b$, $\lambda_{1,1} = 3.832$ is first root of Bessel function $J_1(x)$, $s_0 = -J_0(\lambda_{1,1}) = 0.4028$. The transverse profile (44) has the following properties $R(0) = 1$, $R(1) = 0$, $R'(1) = 0$. At the bunch boundary, the bunch density and its derivative are zero. For such electron bunch we have

$$\hat{T}(\Omega_p) = \sqrt{\pi} e^{-\frac{\Omega_p^2}{4}}, \hat{\tau} = \sqrt{\pi},$$

$$Y_n = \frac{s_0}{s_0+1} \frac{J_1(\lambda_n \rho_b)}{\lambda_n \rho_b} \frac{\lambda_{1,1}^2}{\lambda_{1,1}^2 - \lambda_n^2 \rho_b^2}, \hat{\sigma} = \frac{s_0}{s_0+1}, \rho_b = \frac{r_b}{b}.$$

As a result, for the electric potential of plasma oscillations, we obtain the following expression

$$\Phi_w(\rho, \xi, \tau) = 4Q k_p e^{-\frac{\tau_b^2}{4}} \left[ S_0(\rho) \sin(\tau - \xi) - S(\rho, \xi) \sin \tau \right], \quad (45)$$

where

$$S(\rho, \xi) = \sum_{n=1}^{\infty} \left[ \frac{2 J_1(\lambda_n \rho_b)}{\lambda_n \rho_b} \frac{\lambda_{1,1}^2}{(\lambda_{1,1}^2 - \lambda_n^2 \rho_b^2)(\lambda_n^2 + \rho_p^2)} \times \right. \quad (46)$$

$$\left. \times \frac{J_0(\lambda_n \rho)}{J_1^2(\lambda_n)} e^{-\lambda_n \frac{\xi}{\rho_p}} \right],$$

$$S_0(\rho) = S(\rho, \xi = 0).$$

In relations (45, (46)), the following dimensionless variables and parameters are used

$$\rho = \frac{r}{b}, \xi = k_p z, \tau = \omega_p t, \tau_b = \omega_p t_b, \rho_p = k_p b.$$

The first term in the expression for the electric potential (45) describes a wake plasma wave excited by a relativistic electron bunch in an infinite plasma waveguide. We note that the case of an infinite plasma waveguide was studied in [9, 11, 12]. The second term in this expression describes transition plasma oscillations localized in the vicinity of the input conductive end of the plasma waveguide. The spatial distribution of transition plasma oscillations is described by the function $S(\rho, \xi)$. It follows from the form of this function (46) that the transition plasma oscillations are localized in the region of length $z \approx b$ and their amplitude exponentially decreases with distance from this region.

Let us dwell now briefly on the quasi-static electromagnetic field of a relativistic electron bunch in a semi-infinite plasma waveguide. The quasi-static electromagnetic field of an elementary ring bunch (3) is described by expression (28). Integrating in accordance with formula (8) over the entry times and initial radial coordinates of elementary ring charges, we obtain the following expression for the quasi-static field of an electron bunch in a semi-infinite plasma waveguide

$$E_{rst}(r,z,t) = 2\frac{Q}{\hat{\sigma}\hat{\tau}b} \sum_{n=1}^{\infty} \frac{\omega_{nst}}{v_0} \Pi_n \left[ Z_n^{(-)}(t - z/v_0) - Z_n^{(+)}(t + z/v_0) \right], \quad (47)$$

$$Z_n^{(-)}(t - z/v_0) = \frac{1}{t_b} \int_{-\infty}^{t} T\left(\frac{t_0}{t_b}\right) e^{-\omega_{nst}|t - z/v_0 - t_0|} dt_0, \quad (48)$$

$$Z_n^{(+)}(t + z/v_0) = \frac{1}{t_b} \int_{-\infty}^{t} T\left(\frac{t_0}{t_b}\right) e^{-\omega_{nst}(t + z/v_0 - t_0)} dt_0, \quad (49)$$

$$\Pi_n(r) = \frac{\lambda_n Y_n J_1\left(\lambda_n \frac{r}{b}\right)}{J_1^2(\lambda_n)(\lambda_n^2 + k_p^2 b^2)}.$$

Function (48) describes the distribution of the quasi-static field of a relativistic electron bunch in the longitudinal direction at each moment of time. The similar function (49) describes the same distribution of the field of a positively charged image. The transition quasi-static field of a positively charged image is localized in the vicinity of the input end of the waveguide and quickly decreases with time as the bunch moves away from the end of the plasma waveguide.

Let us turn to the study of a transition electromagnetic pulse excited by a relativistic electron bunch. Using Green's function (37), we obtain the following expression describing the shape of the radiated transition pulse for an electron bunch with arbitrary longitudinal and transverse profiles

$$E_r^{(rad)} = -\frac{\sqrt{2\pi}Q}{s_{eff} t_{eff}} \frac{\rho_b^2}{\rho_p} \sum_{n=1}^{\infty} \left(\frac{\omega_p}{\omega_{nc}}\right)^{3/2} \times$$

$$\times \frac{\lambda_n J_1\left(\lambda_n \frac{r}{b}\right)}{J_1^2(\lambda_n)} Y_n L_n(z,t), \quad (50)$$

where

$$L_n(z,t) = \int_{-\infty}^{t} T\left(\frac{t_0}{t_b}\right) A_n(\varsigma, t - t_0) e^{-i\omega_{nc} \vartheta(z, t-t_0)} dt_0 + c.c., \quad (51)$$

$$A_n(\varsigma, t - t_0) = \frac{f(\varsigma)}{\sqrt{\omega_p(t-t_0)\varepsilon_n(\varsigma)}},$$

$$\vartheta(z, t-t_0) = \sqrt{(t-t_0)^2 - z^2/c^2}.$$

We will consider the moments of time $t \gg t_b$ when the bunch as a whole is completely in the plasma. In this case, the upper limit of integral (51) can be replaced by $+\infty$. In the case of a short electron bunch with a smooth longitudinal profile, for example, Gaussian one (43), the main contribution to the integral $L_n(z, t-t_0)$ will be given by the vicinity of the point $t_0 = 0$. This allows us to expand the phase function $\vartheta(z, t-t_0)$ in the vicinity of this point

$$\vartheta(z, t-t_0) \approx \vartheta(z,t) - t_0 \frac{t}{\vartheta(z,t)}, \vartheta(z,t) = \sqrt{t^2 - z^2/c^2}.$$

As a result, for the integral (51) we obtain the asymptotic representation

$$L_n(z,t) = 2t_b \hat{T}_n(\zeta) A_n(\zeta, t) \cos\left(\omega_{nc} t \sqrt{1-\zeta^2}\right),$$

where $\hat{T}_n(\zeta) \equiv \hat{T}(\Omega_{ns}(\zeta))$, $\Omega_{ns}(\zeta) = \omega_{ns}(\zeta) t_b$,

$$\omega_{ns}(\zeta) = \frac{\omega_{nc}}{\sqrt{1-\zeta^2}}, \zeta = \frac{z}{ct}, \quad (52)$$

$\hat{T}(\Omega_{ns}(\zeta))$ is Fourier amplitude of the function $T(t/t_b)$ at the dimensionless frequency $\Omega_{ns}(\zeta)$,

$$\hat{T}(\Omega_{ns}(\zeta)) = 2\int_0^{\infty} T(\tau_0) \cos(\Omega_{ns}(\zeta)\tau_0) d\tau_0.$$

Accordingly, for a transition electromagnetic pulse, instead of (50), we obtain

$$E_r^{(rad)} = -2\sqrt{\frac{2}{\pi}} \frac{Q}{b^2} \frac{1}{\hat{\tau}\hat{\sigma}\rho_p} \sum_{n=1}^{\infty} \left[ \left(\frac{\omega_p}{\omega_{nc}}\right)^{3/2} \frac{\lambda_n Y_n}{J_1^2(\lambda_n)} J_1\left(\lambda_n \frac{r}{b}\right) \times \right.$$

$$\times \hat{T}_n(\zeta) A_n(\zeta,t) \cos\left(\left(1-\zeta^2\right)^{1/2} \omega_{nc} t\right)\Big], \qquad (53)$$

$$\hat{T}_n(\zeta) \equiv \hat{T}(\Omega_{ns}(\zeta)).$$

For the magnetic field of the transition radiation, we have the following expression

$$H_\varphi^{(rad)} = -2\sqrt{\frac{2}{\pi}} \frac{Q}{b^2} \frac{1}{\hat{\tau}\hat{\sigma}\rho_p} \sum_{n=1}^{\infty}\left[\left(\frac{\omega_p}{\omega_{nc}}\right)^{3/2} \frac{\lambda_n Y_n}{J_1^2(\lambda_n)} J_1\left(\lambda_n \frac{r}{b}\right) \times\right.$$

$$\left.\times \hat{T}_n(\zeta) \frac{\varepsilon_n(\zeta)}{\zeta} A_n(\zeta,t) \cos\left(\omega_{nc} t \sqrt{1-\zeta^2}\right)\right]. \quad (54)$$

The total transition electromagnetic pulse is a discrete superposition of electromagnetic pulses, each of which corresponds to a radial harmonic of the plasma waveguide. It follows from formulas (53), (54) that the structure of the field of an electromagnetic pulse of each radial harmonic excited by a real bunch differs from the pulse of an infinitely thin ring bunch by the presence of form-factor $\hat{T}(\Omega_{ns}(\zeta))$, which in turn is given by a specific longitudinal profile of the electron bunch.

For a bunch with Gaussian longitudinal (43) and model transverse (44) profiles, the expressions for the electromagnetic field of the transition radiation take the form

$$E_r^{(rad)} = -\sqrt{\frac{8}{\pi}} \frac{Q}{b^2} \frac{1}{\rho_p \sqrt{\omega_p t}} \sum_{n=1}^{\infty}\left[\left(\frac{\omega_p}{\omega_{nc}}\right)^{3/2} \frac{2J_1(\lambda_n \rho_b)}{\lambda_n \rho_b} \frac{\lambda_n}{J_1^2(\lambda_n)} \times\right.$$

$$\left.\times \frac{1}{1-\frac{\lambda_n^2}{\lambda_{1,1}^2}\rho_b^2} J_1\left(\lambda_n \frac{r}{b}\right) \frac{g_n(\zeta)}{\varepsilon_n(\zeta)} \cos\left(\omega_{nc} t \sqrt{1-\zeta^2}\right)\right], \quad (55)$$

$$H_\varphi^{(rad)} = -\sqrt{\frac{8}{\pi}} \frac{Q}{b^2} \frac{1}{\rho_p \sqrt{\omega_p t}} \sum_{n=1}^{\infty}\left[\left(\frac{\omega_p}{\omega_{nc}}\right)^{3/2} \frac{2J_1(\lambda_n \rho_b)}{\lambda_n \rho_b} \frac{\lambda_n}{J_1^2(\lambda_n)} \times\right.$$

$$\left.\times \frac{1}{1-\frac{\lambda_n^2}{\lambda_{1,1}^2}\rho_b^2} J_1\left(\lambda_n \frac{r}{b}\right) \frac{g_n(\zeta)}{\zeta} \cos\left(\omega_{nc} t \sqrt{1-\zeta^2}\right)\right], \quad (56)$$

$$g_n(\zeta) = e^{-\frac{\Omega_{nc}^2}{4(1-\zeta^2)}} f(\zeta),$$

$\Omega_{nc} = \omega_{nc} t_b$. We note that at $\omega_p t_b \gg 1$, the amplitudes of all radial harmonics in sums (55), (56) are exponentially small. In the case of a short electron bunch $\omega_p t_b \ll 1$, there are always excited a finite number $n_{max}$ of radial harmonics of the plasma waveguide, the numbers of which satisfy the condition $\lambda_n \approx \pi(n_{max} - 1/4) < b/2ct_b$. The longitudinal profile of the amplitude of electromagnetic pulses corresponding to the radial harmonics in the sums (55), (56) at each moment of time is described mainly by the function $g_n(\zeta)$. This function vanishes at the leading edge $\zeta = 1$ of the transition pulse and in the plane of injection $\zeta = 0$ of the electron bunch and reaches its maximum value at the point $\zeta = \zeta_m$, whose coordinate is the root of the equation

$$\frac{(1-\zeta^2)(2+\zeta^2) - \gamma_0^{-2}(2-\zeta^2)}{1-\zeta^2 + \gamma_0^{-2}\zeta^2} = \frac{\Omega_{nc}^2 \zeta^2}{1-\zeta^2}. \quad (57)$$

In the most interesting case of a short $\Omega_{nc}^2 \ll 1$ relativistic $\gamma_0^{-2} \ll 1$ electron bunch, equation (57) can be simplified and represented in the form

$$\left(1-\zeta^2\right)^2 - \frac{1}{3}\left(\gamma_0^{-2} + \Omega_{nc}^2\right)\left(1-\zeta^2\right) - \frac{1}{3}\gamma_0^{-2}\Omega_{nc}^2 = 0.$$

The positive root of this equation is easily found

$$1 - \zeta_m^2 = \frac{1}{6}\left[\gamma_0^{-2} + \Omega_{nc}^2 + \sqrt{\left(\gamma_0^{-2} + \Omega_{nc}^2\right)^2 + 12\gamma_0^{-2}\Omega_{nc}^2}\right]. \quad (58)$$

In the most interesting case $\Omega_{nc}^2 \gg \gamma_0^{-2}$, from (58) we obtain

$$\zeta_m = 1 - \frac{1}{6}\Omega_{nc}^2 \qquad (59)$$

or

$$z_m = v_m t, \; v_m = c\left(1 - \frac{1}{6}\Omega_{nc}^2\right). \qquad (60)$$

At point (59), the function $g_n(\zeta)$ reaches its maximum value

$$g_n(\zeta_m) = \frac{e^{-3/4} 3^{3/4}}{\Omega_{nc}^{3/2}} \gg 1.$$

The frequency (33) of the wave packet at this point is equal

$$\omega_{ns}(\zeta = \zeta_m) = \sqrt{3}/t_b.$$

The value of this frequency is determined only by the duration of the electron bunch. In the considered case of a short relativistic electron bunch $\Omega_{nc} = \omega_{nc} t_b \ll 1$ it significantly exceeds the cutoff frequency and, accordingly, the plasma frequency.

If the condition $\Omega_{nc}^2 \ll \gamma_0^{-2}$ is satisfied, then from expression (58) implies formula (39) obtained for the case of an infinitely thin ring bunch.

It follows from formula (60) that each wave packet corresponding to of the radial harmonics with number $n$ propagates in the plasma waveguide at its own velocity. Moreover, with increasing of the number of the radial harmonic, the velocity $v_m$ of movement of the pulse maximum amplitude decreases. This leads to the fact that the initial pulse splits with time into a chain of wave packets following one after another. Moreover, at the beginning, the wave packet with the maximum number $n_{max}$ is detached, and then the other packets are detached as their numbers decrease. Full splitting of the initial electromagnetic pulse to a chain of individual pulses occurs in a time

$$t \propto \frac{6}{\lambda_2^2 - \lambda_1^2} \frac{b}{ct_b} \frac{b}{c}$$

or at the distance

$$z \propto \frac{6}{\lambda_2^2 - \lambda_1^2} \frac{b}{ct_b} b$$

from the plane of injection of the electron bunch.

Let us consider the question about the power and full energy of a transition electromagnetic pulse. The pulse power is defined as the energy flow averaged over high-frequency oscillations carried by the transition electromagnetic pulse through the waveguide cross section

$$P = \frac{c}{4\pi} \int_0^b \overline{E_r^{(rad)}(r,\zeta,t) H_\varphi^{(rad)}(r,\zeta,t)} 2\pi r dr . \quad (61)$$

The bar means averaging over fast oscillations with time. Taking into account formulas (53), (54) from this relation we find the following expression for the radiated power

$$P = \frac{cQ^2}{\pi b^4} \frac{1}{(\hat{\tau}\hat{\sigma})^2} \frac{f^2(\zeta)}{\zeta} \frac{1}{t} \sum_{n=1}^\infty \hat{T}_n^2(\zeta) \frac{v_0^2}{\omega_{nc}^3} \frac{\lambda_n^2 Y_n^2}{\varepsilon_n(\zeta) J_1^2(\lambda_n)} . \quad (62)$$

The sum in (62) takes into account the partial contribution to the radiated power of all radial modes of the plasma waveguide. The energy of an electromagnetic pulse is determined by the integral over time

$$W = \int_{z/c}^\infty P(\overline{t},\zeta) d\overline{t} .$$

Performing the integration, we obtain the following expression for the full energy of the transition electromagnetic pulse

$$W = \frac{c}{\pi} \frac{1}{(\hat{\tau}\hat{\sigma})^2} \frac{Q^2}{b^2} \sum_{n=1}^\infty \frac{v_0^2}{\omega_{nc}^3 b^2} \frac{\lambda_n^2 Y_n^2}{J_1^2(\lambda_n)} I_n, \quad (63)$$

where

$$I_n = \int_{z/c}^\infty \frac{dt}{t} \frac{\hat{T}_n^2(\zeta) f^2(\zeta) d\zeta}{\zeta \varepsilon_n(\zeta)} . \quad (64)$$

For a bunch with Gaussian longitudinal (43) and model transverse (44) profiles, the expression for the full energy of the transition electromagnetic pulse (63) can be transformed to the form

$$W = \frac{1}{\pi} \frac{Q^2}{b} \beta_0^2 S_w, \quad (65)$$

$$S_w = \sum_{n=1}^\infty \frac{1}{q_{np}^3} \left[ \frac{2J_1(\lambda_n \rho_b)}{\lambda_n \rho_b} \right]^2 \frac{\lambda_n^2 I_n}{J_1^2(\lambda_n)\left(1 - \frac{\lambda_n^2}{\lambda_{1,1}^2}\rho_b^2\right)^2},$$

$$q_{np} = \sqrt{\rho_p^2 + \lambda_n^2}, \quad \rho_p^2 = \frac{\omega_p^2 b^2}{c^2},$$

$$I_n = \int_{z/c}^\infty \frac{dt}{t} e^{-\frac{\Omega_{nc}^2}{2(1-\zeta^2)}} \frac{\zeta\sqrt{1-\zeta^2}}{\varepsilon_n(\zeta)(1-\beta_0^2\zeta^2)^2} . \quad (66)$$

Let us define the efficiency of the radiator $\eta$ based on transition radiation as the ratio of the radiated energy (63) to the kinetic energy of the bunch $W_{kin} = N_0 mc^2(\gamma_0 - 1)$, where $N_0 = Q/e$ is the number of particles in the bunch. Then for the efficiency we obtain the expression

$$\eta = \frac{1}{\pi} N_0 \frac{r_{cl}}{b} \frac{\beta_0^2}{\gamma_0 - 1} S_w . \quad (67)$$

where $r_{cl} = e^2 / mc^2$ is the classical radius of the electron.

Integral (66) is identically equivalent to the following

$$I_n = e^{-\frac{\Omega_{nc}^2}{2}} \int_0^\infty \frac{e^{-\frac{\Omega_{nc}^2}{2}x^2}(1+x^2)dx}{(1+x^2/\gamma_0^2)^2 (q_{np}^{-2}+x^2)}, \quad (68)$$

In the most interesting case of a short $\Omega_{nc}^2/2 \ll 1$ relativistic $\gamma_0^{-2} \ll 1$ electron bunch, integral (68) can be calculated approximately. Its value depends on the ratio between the above small parameters. If the bunch is so thin that

$$\Omega_{nc}^2 / 2 \ll \gamma_0^{-2}, \quad (69)$$

then the value of integral (68) is approximately equal to

$$I_n = \frac{\pi}{4} \gamma_0 \frac{q_{np}^2(\gamma_0^2 + 2q_{np}\gamma_0 + 1)}{(1+q_{np}\gamma_0)^2} . \quad (70)$$

Since we can also assume that $\gamma_0 \gg q_{np}$, then formula (70) takes on the simple form

$$I_n = \frac{\pi}{4} \gamma_0 .$$

The full energy of the transition electromagnetic pulse is proportional to the relativistic factor of the bunch, and the efficiency of the radiator does not depend on the relativistic factor and is determined only by the number of particles in the bunch and the parameters of the waveguide (waveguide radius and plasma density). We also note that in the considered limiting case (69), the full energy and efficiency do not depend on the duration of the relativistic electron bunch. This limiting case corresponds to the model of an infinitely thin bunch.

In the opposite limiting case

$$\Omega_{nc}^2 / 2 \gg \gamma_0^{-2}, \quad (71)$$

which is more realistic, for the integral $I_n$ we obtain the following expression

$$I_n = \frac{\sqrt{2\pi}}{\Omega_{nc}} e^{-\frac{\Omega_{nc}^2}{2}} + \sqrt{\pi} \frac{q_{np}^2-1}{q_{np}} e^{-\frac{\Omega_{nc}^2}{2}\frac{q_{np}^2-1}{q_{np}^2}} \int_{\Omega_{nc}/\sqrt{2}q_{np}}^\infty e^{-x^2} dx . \quad (72)$$

This asymptotic representation is valid for all values $\Omega_{nc}$ within inequality (71). For the short bunch $\Omega_{nc} \ll q_{np}^{-1} < 1$, expression (72) takes the simple form

$$I_n = \frac{\sqrt{2\pi}}{\Omega_{nc}} . \quad (73)$$

In this case, the energy of the transition pulse does not depend on the relativistic factor of the electron bunch, and, accordingly, the efficiency is inversely proportional $\gamma_0$. It also follows from formula (73) that the energy of the electromagnetic pulse and the efficiency of the beam radiator are inversely proportional to the duration of the electron bunch. With an increase in the duration of the electron bunch (an increase in the size of the radiation source), the degree of coherence of the radiation of electromagnetic waves deteriorates.

## CONCLUSION

In the present work, the process of excitation of a transition electromagnetic field by an electron bunch in a semi-infinite plasma waveguide is investigated. The full electromagnetic field contains three components. First of all, in the volume of the plasma, the electron bunch will excite potential longitudinal plasma oscillations. The field of plasma oscillations includes a wake plasma wave, which has the same structure as in an infinite plasma waveguide, as well as a clot of plasma oscillations localized in the vicinity of the input end of the waveguide. The amplitude of these oscillations at the input end of the waveguide is equal to the amplitude of the wake wave, but is opposite in phase and decreases with distance from the input end of the waveguide into the plasma. The presence of the input end in the plasma waveguide also distorts the picture of the own quasi-static electromagnetic field of the relativistic electron bunch. The currents and charges induced at the input end of the waveguide induce a transition quasi-static electromagnetic field of a positively charged mirror «image» of an electron bunch in the plasma. This field is also localized in the vicinity of the input end of the waveguide and has a pulsed nature. The value of this field decreases quickly as the bunch is removed from the input end into the depth of the plasma. And, finally, the full electromagnetic field contains a transition electromagnetic pulse, excited by an electron bunch during its crossing the plasma boundary. The electromagnetic pulse has a sharp leading edge, duration (longitudinal size) of which is determined, first of all, by the duration of the electron bunch itself. The high-frequency component of the electromagnetic pulse is concentrated in the region of the leading edge. With distance from the leading edge to the injection plane, the frequency of the electromagnetic pulse decreases, and the wavelength, accordingly, increases. During its propagation, the initial pulse is split into a chain of electromagnetic pulses, each of which corresponds to the excited radial harmonic of the plasma waveguide.

## REFERENCES

1. Chen P., Dawson J.M., Huff R., Katsouleas T. Acceleration of electrons by the interaction of a bunched electron beam with a plasma . Phys. Rev. Letters. 1985. Vol.54, No.7. P.692-695.
2. Ruth B.D., Chao A.W., Morton P.L., Wilson P.B. A Plasma wake field accelerator. Particle accelerations. 1985. Vol.17. P.143-155.
3. Katsouleas T. Physical mechanisms in the plasma wake field acceleration. Phys. Rev. A. 1986. Vol.33. No.3. P.2056-2064.
4. Keinigs R., Jones M.E. Two dimensional dynamics of plama wakefield accelerator. Phys. Fluids. 1987. Vol.30, No.1. P.252-263.
5. Kuklin V.M. One-dimensional moving bunches in plasma. Ukr. Phys.J. 1986. Vol.31.No.6. P.853-857.
6. Balakirev V.A., Onishchenko I.N., Fainberg Ya.B. Wake field excitation in plasma by a train of relativistic electron bunches Plasma Physics Reports.1994. Vol.20, No.7. P.606-612.
7. Balakirev V.A., Bliokh Yu.P., Onishchenko I.N., Fainberg Ya.B. Dynamics of the excitation of plasma oscillations by a sequence of bunches of charged particles. Plasma Phys. Reports, 1988. Vol.14. No.2. P.218-225.
8. Balakirev V.A., Sotnikov G.V., Fainberg Ya.B. Modulation of relativistic bunches in a plasma. Plasma Physics Reports. 1996. Vol.22. No.2. P.151-155.
9. Balakirev V.A., Karbushev N.I., Ostrovskii A.O. Theory of Cherenkov Amplifiers and Generators Based on Relativistic Beams. Kiev: «Naukova Dumka». 1993. P.208 .
10. Keinigs R., William P., Jones M.E. A comparison of the dielectric and plasma wake-field accelerators. Phys. Fluids B. 1989. Vol.1. No.1. P.1982-1989.
11. Balakirev V.A., Karas V.I., Tolstoluzskii A.P. Excitation of wake field in a plasma with radial density variation. Plasma Physics Reports. 1997. Vol.23. No.4. P.290-298.
12. Balakirev V.A., Karas V.I. Tolstoluzhsy A.P., Fainberg Ya.B. Excitation of wake fields by a relativistic electron bunch in a radially inhomogeneous plasma. Physics of plasma. 1997. Vol.23, No.4. P.316-324.
13. Balakirev V.A., Onishchenko I.N. Excitation of the transition radiation by a relativistic electron bunch in the plasma half-space. arXiv: 2011.08616.
14. Balakirev V.A., Sidel'nikov G.L. Excitation of electromagnetic pulse by relativistic electron bunch. Technical Physics. 1999. Vol.44. No.10. P.1209-1214.
15. Balakirev V.A., Onishchenko I.N., Sidorenko D.Yu., Sotnikov G.V. Excitation of wake field by a relativistic electron bunch in a semi-infinite dielectric waveguide. Journal of Experimental and Theoretical physics. 2001. Vol.93, No.1. P.33-42.
16. Balakirev V.A., Onishchenko I.N., Sidorenko D.Yu., Sotnikov G.V. Wide-band emission of relativistic electron beam in Excitation of wake field by a relativistic electron bunch in semi-infinite waveguide. Technical Physics. 2002. Vol.72, No.2. P.88-95.
17. Balakirev V.A., Onishchenko I.N., Sidorenko D.Yu., Sotnikov G.V. Charged particles accelerated by wake field in a dielectric resonator with exciting electron beam channel. Technical Phisics Letters. 2003. Vol.29. No.7. P.589-591.18.
18. Balakirev V.A., Gaponenko N.I., Gorban`A.M., Gorozhanin D.V., Egorov A.M., Ermolenko V.V., Lonin Yu.F., Onishchenko I.N. Excitement TEM-horn antenna by impulsive relativistic electron beam. Problems of Atomic Science and Technology. 2000. No.3. Series: Plasma Physics (5). P.118-119.
19. Xinlu Xu, David B. Cesar, Sebastein Corde et al. Generation of terawatt, attosecond pulse from relativistic transition radiation. arXiv: 2007.12736.
20. Karpman V.I. Nonlinear waves in dispersion media. Moscow, «Nauka». 1973. P.176.
21. Olver F.W.J. Introduction to asymptotics and special function. Academic Press, New York and London. 1974.
22. Fedoruk M.V. Asymptotics: Integrals and Series. Moscow, «Nauka». 1987. P.544.